\documentclass[twocolumn,aps,prd,showpacs,superscriptaddress,floats,floatfix,showkeys,notitlepage]{revtex4-1}

\usepackage{lipsum}
\usepackage{graphicx}
\usepackage{subfigure}
\usepackage{palatino}
\usepackage{changes}
\usepackage{hyperref}
\hypersetup{colorlinks=true,linkcolor=blue,urlcolor=blue,citecolor=blue}
\usepackage[toc,page]{appendix}
\usepackage[normalem]{ulem}
\usepackage{adjustbox}
\usepackage{latexsym}
\usepackage{amsmath}
\usepackage{amssymb}
\usepackage{amsfonts}
\usepackage{times}
\usepackage{dcolumn}
\usepackage{bm}
\usepackage{array,tabularx,multirow}

\usepackage{braket}
\usepackage{tensor}
\usepackage{slashed}
\usepackage{booktabs} 
\usepackage{float}
\newcommand{\RNum}[1]{\uppercase\expandafter{\romannumeral #1\relax}}

\begin{document}
\title{Study of relativistic accretion flow in the $f(R)$ theory of gravity}

\author{Akhil Uniyal}
\email{akhil$_$uniyal@iitg.ac.in}
\affiliation{Department of Physics, Indian Institute of Technology Guwahati, Guwahati
781039, Assam, India}

\author{Sayan Chakrabarti}
\email{sayan.chakrabarti@iitg.ac.in}
\affiliation{Department of Physics, Indian Institute of Technology Guwahati, Guwahati
781039, Assam, India}

\author{Santabrata Das}
\email{sbdas@iitg.ac.in}
\affiliation{Department of Physics, Indian Institute of Technology Guwahati, Guwahati
	781039, Assam, India}

\date{\today}

\begin{abstract}
\vskip -0.2 cm	
We present the properties of relativistic, inviscid, low angular momentum, advective accretion flow in a $f(R)$ gravity theory that satisfactorily mimics the asymptotically flat vacuum solutions of the Einstein's equations. With this, we solve the governing equations describing the accretion flow and obtain the global transonic accretion solutions in terms of flow energy (${\cal E}$), angular momentum ($\lambda$) and gravity parameter ($A$) that determines the effect of $f(R)$ gravity. We observe that depending on the model parameters, flow may contain either single or multiple critical points. We separate the effective domain of the parameter space in $\lambda-{\cal E}$ plane that admits accretion solutions possessing multiple critical points and observe that solution of this kind continues to form for wide range of the flow parameters. We examine the modification of the parameter space and reveal that it gradually shrinks with the decrease of $A$, and ultimately disappears for $A=-2.34$. Finally, we calculate the disk luminosity ($L$) considering bremsstrahlung emission process and find that global accretion solutions passing through the inner critical point are more luminous compared to the outer critical point solutions.
\end{abstract}

\pacs{-----------------}

\maketitle

\date{\today}

\section{Introduction}\label{sec:intro}

Accretion process around the black hole is considered as the principle source to power the astrophysical objects, namely micro quasars, active galactic nuclei and quasars \cite{Frank-etal2002}. As the black holes classically do not emit anything, not even electromagnetic radiations, the study of the accretion flow in black hole environment ought to provide the intrinsic imprints of the central engine. Meanwhile, numerous works have been performed on the study of accretion processes near the black hole using different physical conditions \cite[and references therein]{Park:1998wp,Yuan:2014gma,Abramowicz-Fragile2013,Font-2000}. However, only a handful of works were reported in the context of hydrodynamical aspects of accretion flow in modified gravity framework. In particular, major efforts were given in studying the particle dynamics around braneworld black holes \cite{Pun:2008ua,Heydari-Fard:2010agr}, slowly rotating black holes in dynamical Chern-Simons modified gravity \cite{Harko:2009kj}, and black holes in Ho\u{r}ava gravity \cite{Harko:2010ua,Harko:2009rp}, to name a few. In addition, accretion physics has also been studied in exotic backgrounds where the central object is a  boson star \cite{Torres:2002td,Guzman:2005bs}, wormhole \cite{Harko:2009xf}, gravastar \cite{Harko:2009gc}, quark star \cite{Harko:2009ysn}, and even a naked singularity \cite{Joshi:2013dva,Kovacs:2010xm}.

Indeed, black holes are one of the most simple compact objects with the strongest gravitational fields around them. They are ubiquitous in the relativistic theory of gravity. Therefore, testing the theory of gravity around such strong-field regime is one of the ways to understand the theory properly. Towards this, the general theory of relativity (GR) has been tested successfully under various different situations \cite{Will:2014kxa}. However, GR also has its own limitations and drawbacks. The theory fails to explain the initial singularity in the cosmic history of the Universe, existence of dark matter and dark energy, and the singularity at the centre of the black holes. Among the above mentioned facts, the acceleration of the Universe, which was discovered by using type Ia supernovae \cite{SupernovaSearchTeam:2001qse} (see \cite{Albrecht:2006um} for more details), still lacks a satisfactory explanation. This discovery, despite having enormous implications in the field of cosmology, also has great significance in fundamental physics. With the help of WMAP and Planck results, it can be concluded that if GR is the correct theory of gravity, then approximately $68\%$ of the energy content of our Universe is not dark or luminous matter, instead is a mysterious form of energy, known as the dark energy. On the other hand, while the baryonic matter content is $5\%$, the dark matter content of the Universe is approximately $27\%$. Therefore, it perhaps will not be an exaggeration to state that the present astrophysical and cosmological models are facing two crucial issues, namely the dark energy problem (see \cite {Brax:2017idh} for a recent review), and the dark matter problem \cite{Bertone:2016nfn}, respectively. 

So far, three main classes of models were proposed to understand the origin of cosmic acceleration, which are namely the introduction of a cosmological constant $\Lambda$ \cite{Padmanabhan:2002ji}, different models of the dark energy \cite {Brax:2017idh} and the modification of gravity. Interestingly, the introduction of cosmological constant in explaining the acceleration of the Universe requires an extreme fine tuning, which seems very unlikely in a realistic scenario. The second class $i.e.$, the models of dark energy, within the context of GR, requires the introduction of fluids with an equation of state (EoS) $P=-\epsilon$, where $P$ and $\epsilon$ are the pressure and energy density of the fluid, and EoS plays a crucial role late in the matter dominated era. Needless to mention that many such models were proposed, however most of them were not physically convincing, and some are even problematic from the fine tuning perspective. The third one, which is the motivating force behind this work, is the modification of gravity \cite[and references therein]{Nojiri:2006ri,Sotiriou:2008rp} instead of introducing exotic matters in the theory. 

One set of such models where the Ricci scalar ($R$) in the Einstein-Hilbert action is replaced by an arbitrary analytic function of $R$, known as the $f(R)$ gravity models, have been extensively studied in both astrophysical as well as cosmological contexts \cite{Capozziello:2003gx, Capozziello:2008qc,Nojiri:2003rz,Nojiri:2003ft,Nojiri:2006gh,Nojiri:2006ri,Sotiriou:2008rp,Carroll:2003wy}. One of the immediate implications of such models was to examine the explanation of cosmic acceleration \cite{Dolgov:2003px}. Indeed, every single form for $f(R)$ may give rise to a new model of gravity, but it is crucial that these theories are getting tested against observations. These observations need not only be cosmological in character, it can also be astrophysical, or can be tests like perihelion shift of Mercury or bending of light. Considering all these, we infer that the study of accretion process around black holes in modified gravity can provide a path way to test the gravity theories against observations. Meanwhile, some efforts were already given in studying the accretion physics in the black hole backgrounds of such $f(R)$ gravity theories where particle dynamics \cite{Pun:2008ae,Soroushfar:2020kgb,Perez:2012bx} around the black hole was considered. 

With this, we mention the motivation of this work which is two folds in nature. The first one is to study the accretion process in a $f(R)$ gravity background which is asymptotically flat and the remaining one is to perform a complete hydrodynamical study of the accretion process in $f(R)$ modified gravity scenario. It is worth mentioning at this point, that having an asymptotically flat vacuum solution in the $f(R)$ gravity is not straightforward, instead it involves big challenge. 

The vacuum solution of $f(R)$ gravity for a static, spherically symmetric space-time were reported by various group of researchers \cite{Multamaki:2006zb,Capozziello:2007wc,Capozziello:2012iea,Shojai:2011yq}. The axisymmetric solutions were also obtained for such theories \cite{Capozziello:2009jg,Gutierrez-Pineres:2012obj}. However, for a large class of spherically symmetric or axisymmetric models, either the Schwarzschild-de Sitter metric turns out to be an exact solution of the field equations, or the metrics have some diverging terms for which they can not be reduced to the Minkowski form in the asymptotic limit. Meanwhile, the properties of the accretion flow in the $f(R)$ gravity with constant Ricci scalar \cite{Perez:2012bx} were studied in such scenarios where asymptotic flatness is not achieved. Very few works have focused on the direction where exact asymptotically flat vacuum solutions were obtained for different forms of the $f(R)$ \cite{Kalita:2019xjq,Nashed:2020mnp}. Accordingly, we study one such asymptotically flat $f(R)$ gravity model \cite{Kalita:2019xjq} in the context of accretion processes around black holes. At the same time, we also note that a complete relativistic hydrodynamical study of the accretion phenomena in the black hole background is also lacking in the literature, with only a few exceptions \cite{Dihingia:2018tlr,Dihingia:2019zos,Dihingia:2020xxc,Sen:2022hzo,Patra-etal2022}. Accordingly, we plan to fill up this gap by considering hydrodynamic description of the accretion flow in modified gravity backgrounds.  We consider a low angular momentum, inviscid accretion disk and perform a fully relativistic hydrodynamical treatment in the above mentioned modified gravity background. In doing so, we assume relativistic equations of state (EoS) \cite{Chattopadhyay:2008xd,Dihingia:2019xdx} and carry out the critical point analysis to obtain and classify the critical points. Using the critical point properties, we solve the governing equations to obtain the transonic global accretion solutions in terms of the input parameters of the theory, such as energy ($\cal E$), angular momentum ($\lambda$), and gravity parameters ($A$) that regulates the gravity effect. Afterwards, we study the parameter space in $\lambda-{\cal E}$ plane, where the boundary of the parameter space indicates the ranges of parameters that separate accretion solutions containing multiple critical points. Finally, considering free-free emission, we estimate the disk luminosity corresponding to the global accretion solution and study its variation with input parameters. 

The paper is organized as follows: in \S\RNum{2}, we discuss the modified gravity model that is used throughout the paper to study the accretion disk properties. In \S \RNum{3}, we present the model formalism including assumptions and governing equations. In \S \RNum{4}, we discuss the accretion solutions in modified gravity including all results. Finally, we conclude the paper by giving a brief summary and outlook in \S \RNum{5}. 

\section{Description of Modified Gravity}

\label{sec:formal}

The energy-momentum tensor for the non-dissipative fluid composed by ions and electrons is described by the following equation:
\begin{equation} \label{e1}
T^{\mu \nu}=(e + P) u^\mu u^\nu + P g^{\mu \nu},
\end{equation}
where $e$, $P$ and $u^\mu$ are the energy density, pressure and four velocity vector, respectively. Here, $\mu$ and $\nu$ represent component of the coordinates ($t,r,\theta,\phi$), and in this work, $g^{\mu \nu}$ refers the metric component which is described as $g^{\mu \nu}=(g^{tt},g^{rr},g^{\theta \theta},g^{\phi \phi})$. In order to understand the effect of modified gravity, we consider asymptotically flat spherically symmetric metric $g_{\mu \nu} = {\rm diag}(-s(r),p(r),r^2,r^2 \sin^2{\theta})$, where metric component $s(r)$ and $p(r)$ are the functions of radial coordinate $r$ only. In this work, we adopt a model of modified gravity which we describe below.

Following \cite{Kalita:2019xjq}, we consider $F(r)=df(R)/dR=1+A/r$, which yields $F(r) \to 1$ for $r \to \infty$. Here, $A$ is a parameter (hereafter gravity parameter) with which gravity is being modified. Indeed, the above choice of $F(r)$ satisfies the criteria of asymptotic flat vacuum solutions. Accordingly, the metric components are given by,
\begin{equation} \label{e2}
\begin{split}
	s(r)=1-\frac{2M_S}{r}-\frac{A(-6M_S+A)}{2r^2}+\frac{A^2(-66M_S+13A)}{20r^3}\\
	-\frac{A^3(-156M_S+31A)}{48r^4}+\frac{3A^4(-57M_S+11A)}{56r^5}\\
	-\frac{A^5(-360M_S+67A)}{128r^6}+ \hdots,
\end{split}
\end{equation}
and
\begin{equation}\label{e3}
	p(r)=\frac{X(r)}{s(r)},
\end{equation}
where $X(r)=\frac{16r^4}{(A+2r)^4}$. Accordingly, $f(R)$ is given by,
\begin{equation} \label{e4}
f(R(r))=R+K_{1}R^{5/4}+K_{2}R^{3/2}+ \hdots,
\end{equation}
where 
$K_{1}=\frac{12}{5.3^{5/4}} \frac{A^{3/4}}{(A-2)^{1/4}}$ and $K_{2}=\frac{1}{60 \sqrt{3}} \frac{A^{3/2} (A-12)}{(A-2)^{3/2}}$.

\section{Model formalism}

We begin with a steady, axisymmetric, inviscid, low angular momentum, advective accretion flow that resides at the equatorial plane of the central source \cite{Chakrabarti:1989,Das-etal2001a,Dihingia:2019xdx}. In studying the properties of the accretion flow, we adopt the theory of modified gravity and present the model equations and critical point analysis in the subsequent sub-sections.

\subsection{Governing Equations for accretion flow}

The conservation of the energy-momentum tensor and mass flux provide the relevant hydrodynamical equations that describe the flow and are given by,
\begin{equation}\label{e5}
T^{\mu \nu}_{;\nu}=0,\qquad {\rm and} \qquad (\rho u^\nu)_{;\nu}=0,
\end{equation}
where $\rho$ refers the local mass density of the flow. The timelike velocity field satisfies the condition $u_\mu u^\mu=-1$. We take the projection of the conservation equation on the spatial hyper-surface and get the Euler equation by using the projection operator $h^{ \alpha}_\mu=\delta^{\alpha}_\mu+u^{ \alpha} u_\mu$ as,
\begin{equation}\label{e6}
h^{ \alpha}_\mu T^{\mu \nu}_{;\nu}=( e + P ) u^\nu u^{ \alpha}_{;\nu}+(g^{{ \alpha} \nu} + u^{ \alpha} u^\nu) P_{,\nu}=0,
\end{equation}
Here, projection operator satisfies the condition $h^{ \alpha}_\mu u^\mu=0$, confirming that the four velocity and projection vectors are orthogonal to each other. Projecting the conservation equation along the direction of the four velocity, we obtain the first law of thermodynamics as,
\begin{equation}\label{e7}
u_\mu T^{\mu \nu}_{;\nu}=u^\mu \left[ \left(\frac{ e + P}{\rho} \right) \rho_{,\mu}-e_{,\mu} \right]=0.
\end{equation}

In this work, we consider the structure of the accretion disk to be geometrically thin and hence, the flow generally lies around the disk equatorial plane with $\theta =\pi/2$. This leads to $u^\theta=0$. Next, we define the three radial velocity of the fluid in the co-rotating frame as  
$v=\gamma_\phi^2 v_r^2$, where, $v_r^2=u^ru_r/(-u^tu_t)$, the azimuthal
Lorentz factor $\gamma_\phi^2=1/(1-v_\phi^2)$, the radial Lorentz factor $\gamma_{v}^2=1/(1-v^2)$ and $v_\phi^2=u^\phi u_\phi/(-u^tu_t)$. Using these definitions of velocities in equation \eqref{e6}, we obtain the radial momentum equation for ${ \alpha}=r$ as
\begin{equation}\label{e8}
v\gamma_v^2\frac{dv}{dr}+\frac{1}{h\rho}\frac{dp}{dr}+\frac{d\Phi_e^{\rm eff}}{dr}=0,
\end{equation}
where $h~[=(e + P)/\rho]$ is the specific enthalpy, and $\Phi_e^{\rm eff}$ represents the effective potential in the disk equatorial plane and is given by,
\begin{equation} \label{e9}
\Phi_e^{\rm eff}=1+\frac{1}{2}\ln{\Phi}; \qquad\qquad \Phi=\frac{-g_{tt}g_{\phi \phi}}{(g_{\phi \phi}+\lambda^2 g_{tt})}.
\end{equation}

The adopted space-time under consideration is stationary and axisymmetric, and hence, there exits mutually perpendicular two killing vectors which allow us to construct two conserve quantities along the direction of flow motion which are given by,
\begin{equation} \label{e10}
		hu_\phi = {\cal L}~({\rm constant});
		\qquad
		-h u_t = {\cal E} ~({\rm constant}),
\end{equation}
where ${\cal E}$ is known as Bernoulli constant (equivalently the specific energy) of the flow and $\cal{L}$ corresponds to the angular momentum conservation. Note that index $\alpha=t,\phi$ in turn renders conserve energy and angular momentum, respectively (see equation \eqref{e10}). We express the specific angular momentum of the flow as $\lambda=\mathcal{L}/\mathcal E = -u_\phi/u_t$, which is also a conserved quantity.

\begin{figure}
	\centering
	\includegraphics[width=\columnwidth]{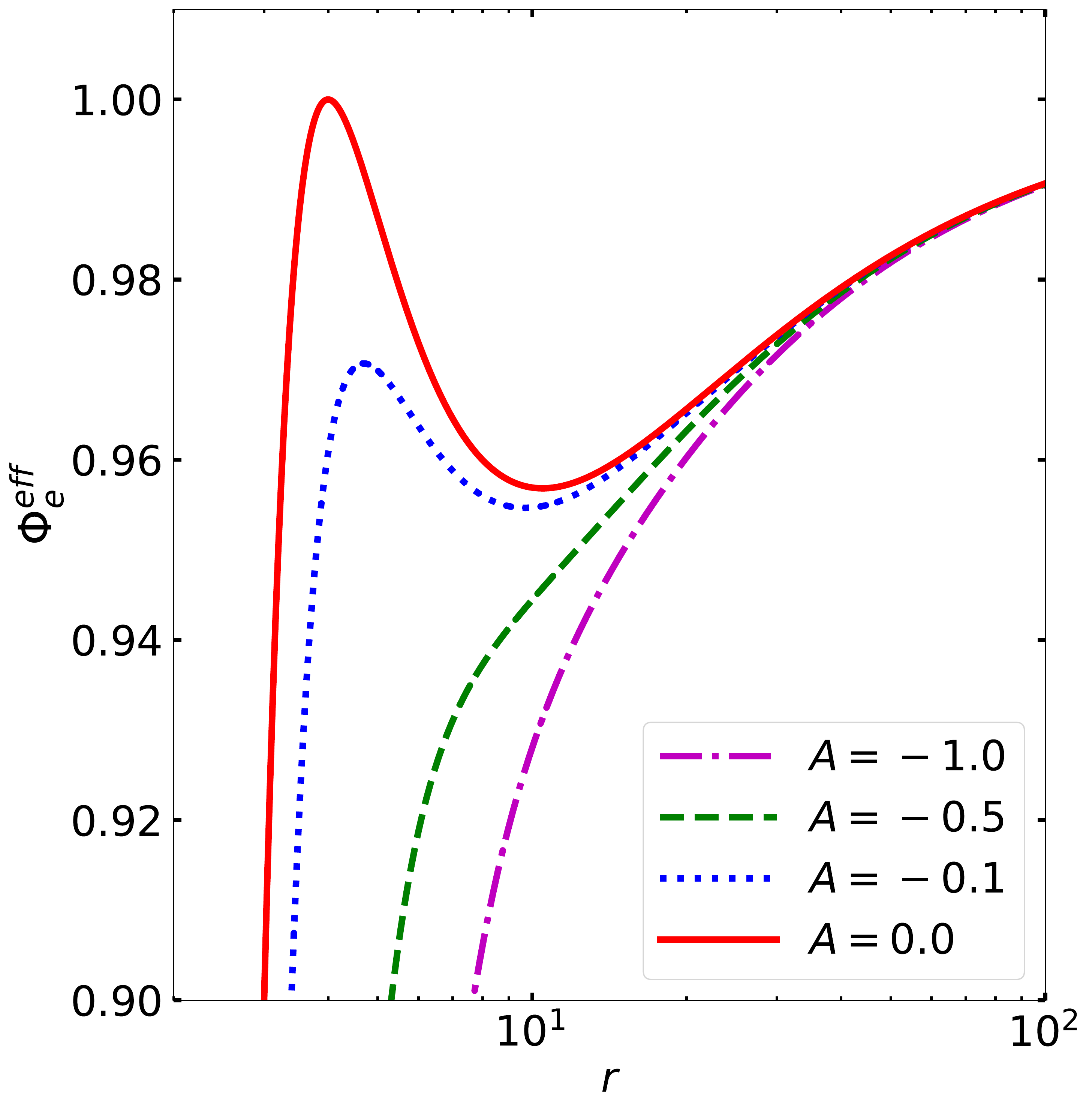}
	\caption{Variation of the effective potential ($\Phi^{\rm eff}_{e}$) as a function of radial coordinate ($r$). Results plotted using dot-dashed (purple), dashed (green), dotted (blue) and solid (red) curves are for $A=-1.0, -0.5, -0.1$, and $0$, respectively. See text for details.
	} 
	\label{Phi-r}
\end{figure}

In order to examine the effect of modified gravity, in Fig. \ref{Phi-r}, we present the variation of the effective potential ($\Phi^{\rm eff}_{e}$) as function of the radial coordinate ($r$) for flows having angular momentum $\lambda=4.0$. Here, dot-dashed (purple), dashed (green), dotted (blue) and solid (red) curves are for $A=-1.0, -0.5, -0.1$, and $0$, respectively. We observe that $\Phi^{\rm eff}_{e}$ tends to merge with the Schwarzschild  potential ($A = 0$) at large $r$ indicating the asymptotically flat vacuum solutions. In addition, we find that when $A$ is decreased, the existence of the local maxima of $\Phi^{\rm eff}_{e}$ near the horizon gradually diminishes. This clearly indicates that for a fixed $\lambda$, modification of gravity eventually weakens the potential barrier.

\begin{figure}
	\centering
	\includegraphics[width=\columnwidth]{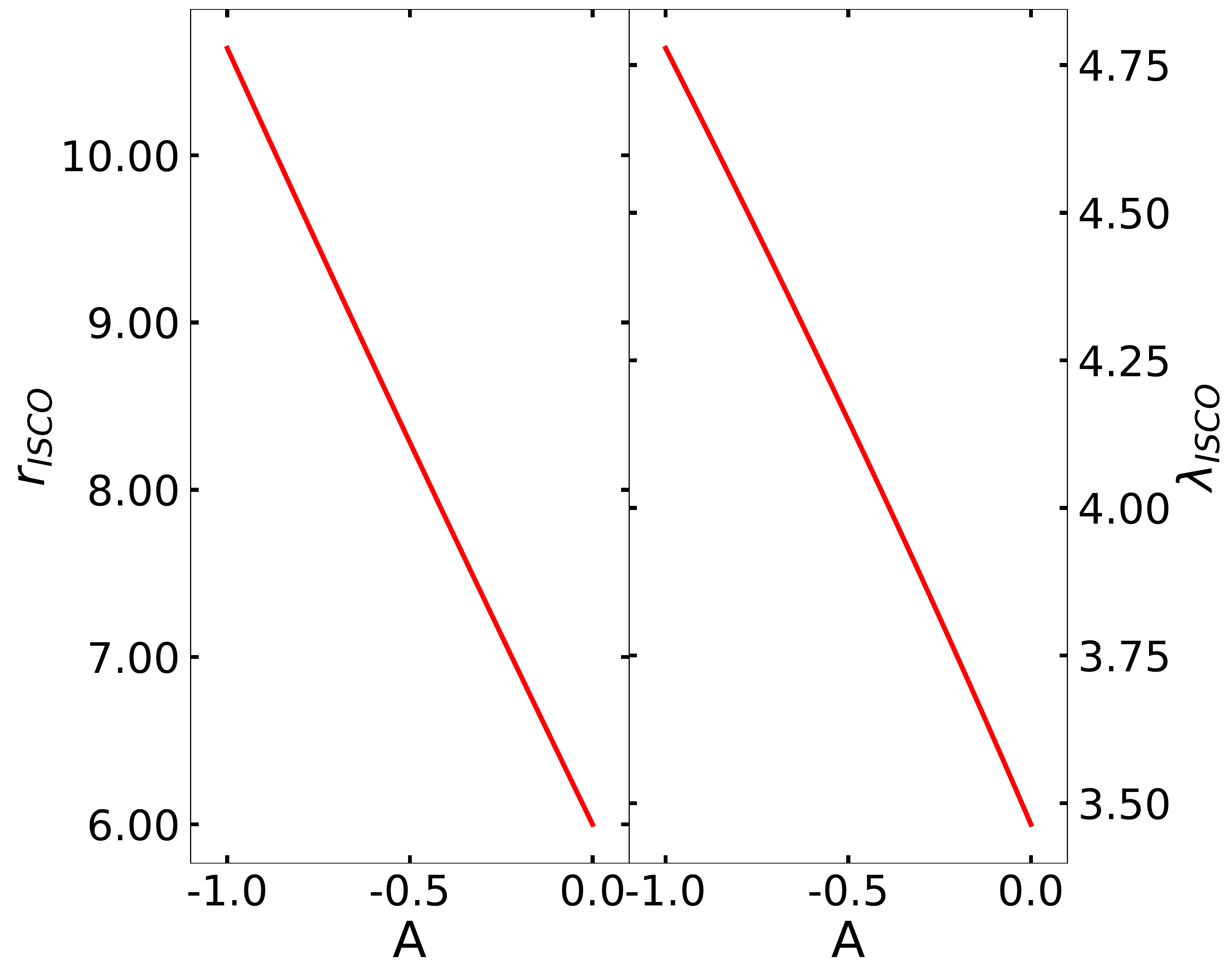}
	\caption{Variation of (a) inner stable circular orbit ($r_{\rm ISCO}$) and (b) the flow angular momentum ($\lambda_{\rm ISCO}$) at $r_{\rm ISCO}$ as function of $A$.  See text for details.
	} 
	\label{isco-r}
\end{figure}

We also calculate the radius of the inner most circular orbit ($r_{\rm ISCO}$) and the corresponding angular momentum ($\lambda_{\rm ISCO}$) at $r_{\rm ISCO}$ and plot them as function of the gravity parameter $A$ in Fig. \ref{isco-r}. We observe that $r_{\rm ISCO}$ recedes away from the horizon as $A$ is decreased. Similarly, $\lambda_{\rm ISCO}$ is also seen to increase with the decrease of $A$. Hence, it is evident that in the modified gravity framework, flow with relatively higher angular momentum can smoothly cross the event horizon in comparison with the Schwarzschild black hole where $A=0$.

The continuity equation (second part of the equations \eqref{e5}) is expressed in terms of the mass accretion rate (${\dot M}$), which is a constant of motion and is given by,
\begin{equation}\label{e11}
	\dot{M}=-4\pi ru^r \rho H,
\end{equation}
where $H$ is the local half-thickness of the accretion disk. In this work, $H$ is calculated assuming the flow to be in hydrostatic equilibrium in the vertical direction and is given by \cite{Riffert:1995,Peitz:1996jp},
\begin{equation} \label{e12}
	H=\sqrt{\frac{Pr^3}{\rho \gamma_\phi^2}}.
\end{equation}

The governing flow equations are closed with an equation of state (EoS) that relates pressure ($P$), density ($\rho$), and internal energy ($e$). Because of the black hole's strong gravity, the accretion flow is expected to be relativistic in nature at the vicinity of the horizon. Hence, following Chattopadhyay and Ryu \cite{Chattopadhyay:2008xd}, we consider an EoS for relativistic flow as,
\begin{equation}\label{e13}
	e=\frac{\rho f}{\left( 1+ \frac{m_p}{m_e}\right)},
\end{equation}
with
$$
f=\left[ 1+\Theta \left( \frac{9\Theta +3}{3\Theta +2} \right) \right] + \left[ \frac{m_p}{m_e} +\Theta \left( \frac{9\Theta m_e + 3 m_p}{3\Theta m_e + 2 m_p}\right) \right],
$$
where $m_e$ and $m_p$ are the mass of the electron and ion, $\Theta~(=k_{\rm B} T/m_e c^2)$ refers to the dimensionless temperature and $k_{\rm B}$ is the Boltzmann's constant. Using the relativistic EoS, we delineate the polytropic index ($N$), adiabatic index ($\Gamma$) and sound speed ($a_s$) as,
\begin{align}\label{e14}
	N = \frac{1}{2} \frac{df}{d\Theta}, \quad \Gamma = 1+\frac{1}{N}, \quad a_s^2 = \frac{\Gamma P}{e+P}=\frac{2\Gamma \Theta}{f+2\Theta}.
\end{align}

Using equation \eqref{e13}, we integrate equation \eqref{e7} to express the flow density ($\rho$) as a function of temperature ($\Theta$) as,
\begin{equation} \label{e15}
	\rho= \mathcal{K} \theta^{3/2} (3\theta+2)^{3/4} (3\theta+2/\chi)^{3/4} \exp \left[ \frac{m_e(f-1)-m_p}{2\Theta m_e}\right],
\end{equation}
where constant $\mathcal{K}$ is the measure of entropy. Subsequently, we define the entropy accretion rate $\dot{\mathcal{M}}$ \cite{Chattopadhyay:2016kcz} as
\begin{equation} \label{e16}
	\begin{split}
		\dot{\mathcal{M}} = \frac{\dot{M}}{4\pi \mathcal{K}} = \theta^{3/2} (3\theta+2)^{3/4} (3\theta+2/\chi)^{3/4} \\
		\times \exp \left[ \frac{m_e(f-1)-m_p}{2\Theta m_e}\right] Hru^r,
	\end{split}
\end{equation}
It is noteworthy that for a non-dissipative flow, $\dot{\mathcal{M}}$ remains conserved all throughout the flow \cite{Dihingia:2020xxc}.

\subsection{Wind Equation}

Now, using equations \eqref{e8}, \eqref{e11} and \eqref{e13}, we calculate the radial derivative of the flow velocity in the form of wind equation as,
\begin{equation} \label{e17}
\frac{dv}{dr}=\frac{\mathcal{N}}{\mathcal{D}},
\end{equation}
where the numerator $\mathcal{N}$ is given by,
\begin{equation} \label{e18}
\mathcal{N}=\frac{2a_s^2}{\Gamma+1} \left [\frac{5}{2r}-\frac{1}{2 p(r)}\frac{d p(r)}{dr}-\frac{1}{2\gamma_\phi^2}\frac{d\gamma_\phi^2}{dr} \right]-\frac{d\Phi^{\rm eff}}{dr}\\
\end{equation}
and the denominator $\mathcal{D}$ is given by,
\begin{equation} \label{e19}
\mathcal{D}=\gamma_v^2 \left(v-\frac{2a_s^2}{v(\Gamma+1)} \right).
\end{equation}

Further, using equations \eqref{e6}, \eqref{e11} and \eqref{e14}, we obtain the gradient of temperature as,

\begin{equation} \label{e20}
\begin{split}
\frac{d\Theta}{dr}=\frac{-2\Theta}{2N+1} \left [\frac{\gamma_v^2}{v} \frac{dv}{dr}+\frac{5}{2r}-\frac{1}{2 p(r)}\frac{dp(r)}{dr}-\frac{1}{2\gamma_\phi^2}\frac{d\gamma_\phi^2}{dr} \right].
\end{split}
\end{equation}

\subsection{Critical point analysis}\label{CPA}

\begin{figure}
	\centering
	\includegraphics[width=\columnwidth]{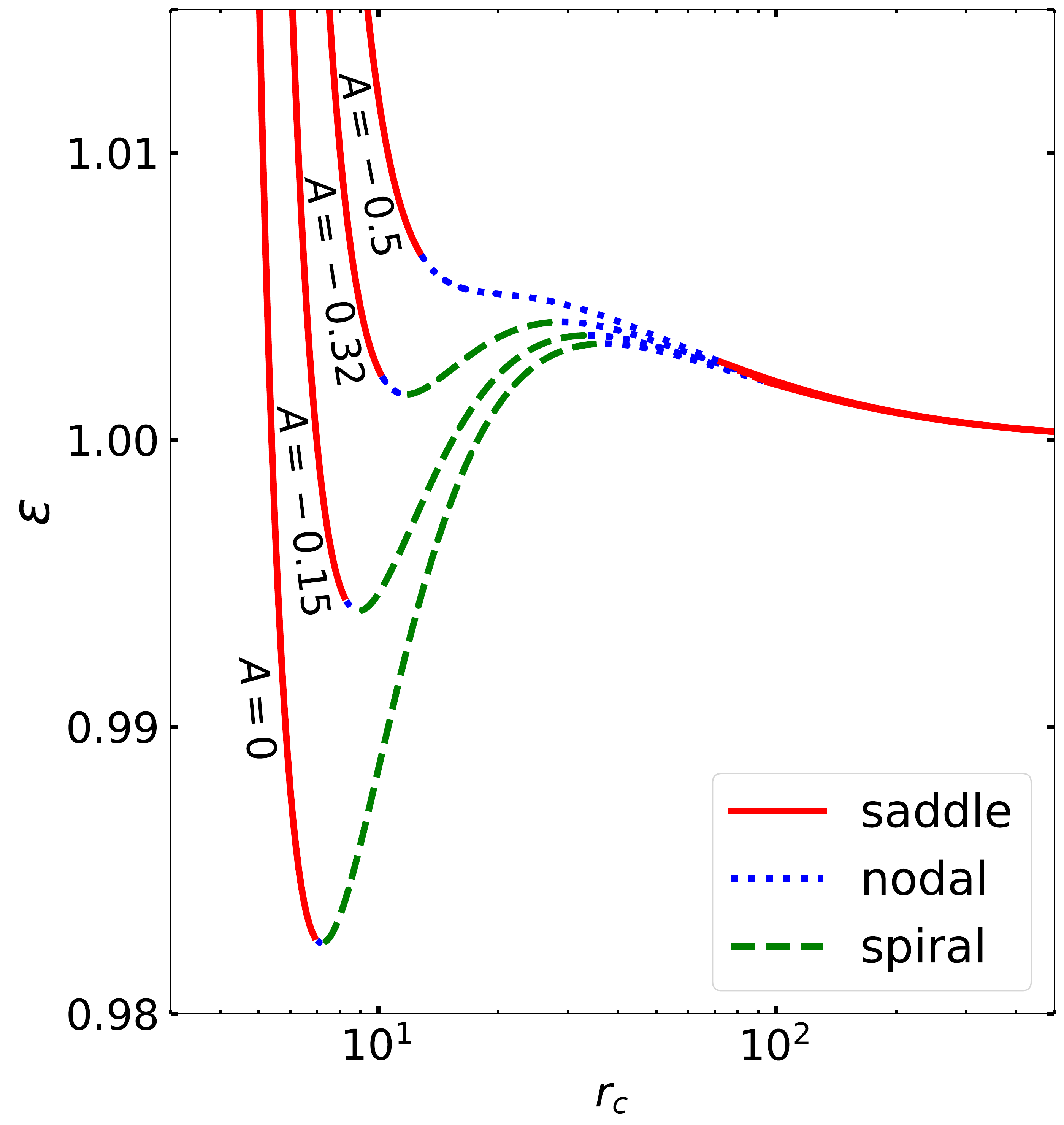}
	\caption{Variation of the specific energy ($\cal E$) as function of critical points ($r_c$). Here, we choose $\lambda=3.2$ and mark the gravity parameter $A=0, -0.15,-0.32$ and $-0.50$. Solid (red), dotted (blue) and dashed (green) parts of a given curve refer saddle, nodal and spiral critical points. See text for details.
	} 
	\label{E-r}
\end{figure}

\begin{figure}
	\centering
	\includegraphics[width=\columnwidth]{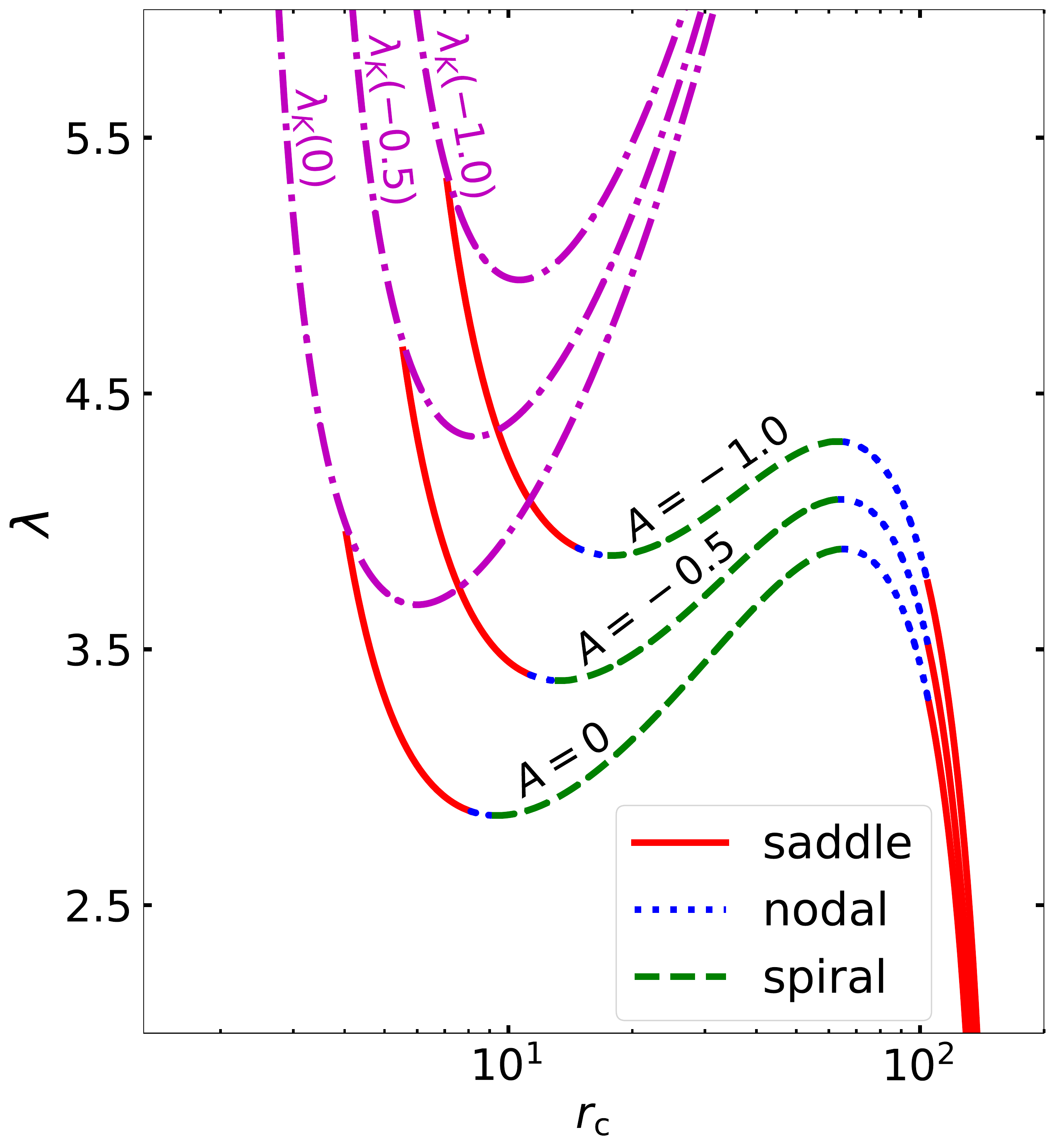}
	\caption{Variation of the specific angular momentum ($\lambda$) as function of critical point location ($r_c$) for a set of gravity parameters ($A$). Obtained results are presented in using solid (red), dotted (blue) and dashed (green) segments which denote saddle, nodal and spiral type critical points. Here, we choose ${\cal E}=1.0018$ and gravity parameters ($A$) are marked. The Keplerian angular momentum profile ($\lambda_K (A)$) is represented by the dot-dashed curve (purple). See text for the details.
	} 
	\label{L-r}
\end{figure}

Over the course of accretion on to a black hole, infalling matter starts its journey with sub-sonic velocity ($v \to 0$) from the outer edge of the disk ($r_{\rm edge}$). Because of the black hole's strong gravity, radial velocity gradually increases as the flow moves towards the horizon  ($r_h$) and eventually makes sonic state transition to become super-sonic after smoothly crossing the critical point ($r_c$). This is inevitable as the flow must satisfy the `supersonic' inner boundary condition at the horizon. Note that, at $r_c$, both numerator and denominator tend to vanish simultaneously ($i.e.$ ${\cal N}={\cal D}=0$) and hence, the radial velocity gradient takes $dv/dr \to 0/0$ form. Since flow remains smooth along the streamline, $dv/dr$ must be real and finite all throughout. Hence, we compute the $(dv/dr)_{r_c}$ by applying the l$'$H\^{o}pital's rule. In reality, $(dv/dr)_{r_c}$ assumes two values: one for accretion and other for wind branch. Depending on the $(dv/dr)_{r_c}$ values, critical points are classified in three categories. When critical points are saddle type, both values of $(dv/dr)_{r_c}$ are real and of opposite sign. For nodal type critical points, the values of $(dv/dr)_{r_c}$ are real and of same sign, whereas for spiral type critical point, $(dv/dr)_{r_c}$ becomes imaginary \cite[and references therein]{Chakrabarti:1989,Das-etal2001a,Das-2007}. It is noteworthy to mention that in the astrophysical context, accretion flow can pass through the saddle critical point only.

In order to evaluate the critical point location, we solve the second part of the equation \eqref{e10} by supplying the input parameters, namely energy ($\mathcal E$), angular momentum ($\lambda$) and gravity parameter ($A$). Depending on the set of chosen parameters ($\cal E$, $\lambda$, $A$), accretion flow may possess single or multiple critical points. Hence, while examining the transonic nature of the flow, we calculate the flow energy ($\cal E$) as a function of the critical point location ($r_c$) for different $A$ parameters. Accordingly, in Fig. \ref{E-r}, we present the obtained results, where we choose $\lambda=3.2$ and vary the gravity parameter as $A=0, -0.15, -0.32$ and $-0.50$, respectively. The solid (red), dotted (green), and dashed (green) segments in each curve denote saddle, nodal and spiral type critical points. It is evident that the critical points occur in sequential order as saddle-nodal-spiral-nodal-saddle-nodal with the increase of $r_c$. Note that when the saddle type critical point forms near the horizon, it is usually called as inner critical point ($r_c \equiv r_{\rm in}$), whereas the same located at far away from the horizon is referred as outer critical point ($r_c \equiv r_{\rm out}$). Further, we observe that flow possesses multiple saddle type critical points for a range of $\cal E$ and $A$ parameter, which is essential for shock transition (see \S IV) \cite[]{Chakrabarti:1989, Das-etal2001a,Dihingia:2018tlr}. Needless to mention that $A=0$ renders the results identical to the Schwarzschild case.

We continue the investigation of the transonic behaviour of the flow and in Fig. \ref{L-r}, we depict the variation of the angular momentum ($\lambda$) as function of the critical point location ($r_c$) for a set of gravity parameters ($A$). Here, we choose ${\cal E}=1.0018$. As before, the solid (red), dotted (blue) and dashed (green) parts of the curve denote saddle, nodal and spiral type critical points, respectively. Setting $d\Phi^{\rm eff}_e/dr=0$, we obtain the Keplerian angular momentum distribution $\left[ \lambda_{K}(A) \right]$, which is depicted using dot-dashed curve (purple). Figure evidently indicates that $\lambda$ remains sub-Keplerian all throughout the disk between the horizon ($r_h$) and the outer edge of the disk ($r_{\rm edge}$).

\begin{figure}
	\centering
	\includegraphics[width=\columnwidth]{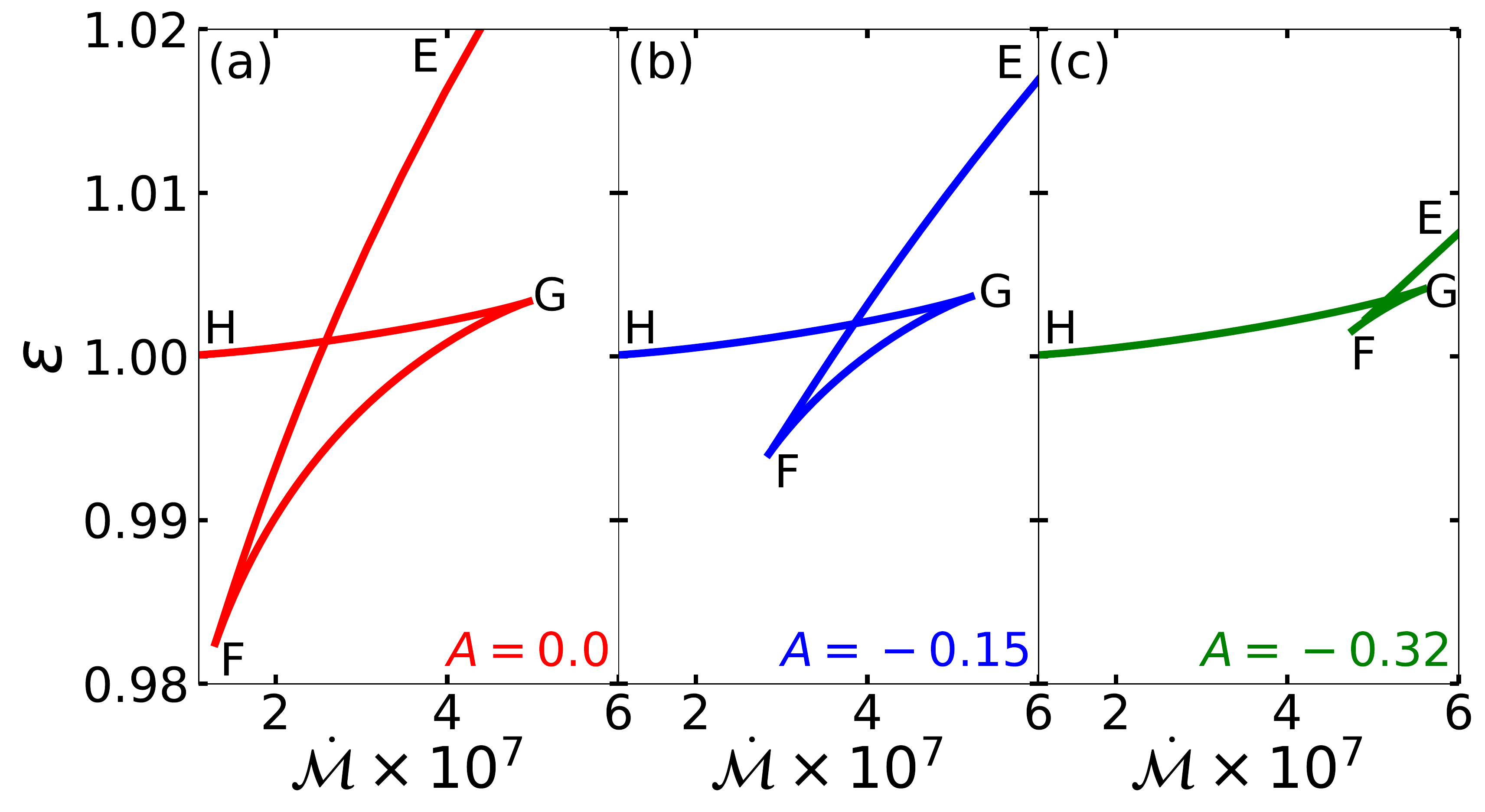}
	\caption{Variation of energy ($\cal E$) with the accretion rate ($\dot{\cal M}$) that renders critical points. Here, we set angular momentum $\lambda=3.2$ and vary gravity parameter as (a) $A=0$, (b) $-0.15$, and $-0.32$, respectively. See text for the details.
	} 
	\label{E-mdot}
\end{figure}

For the purpose of completeness of the critical point analysis, we examine the variation of ${\cal E}$ as a function of ${\dot {\cal M}}$ for
a given set of $\lambda$ and $A$ for all possible accretion solutions. Fig. \ref{E-mdot} shows such plot when $\lambda=3.2$ is chosen and $A$ is varied as $0$, $-0.15$ and $-0.32$ in panel (a), (b) and (c), respectively. In general, a transonic flow with initial flow parameters that lies on the branch EF passes through the inner saddle type critical points and a flow with parameters belonging to the branch GH has the outer saddle type critical  point. The parameters chosen from FG branch give the spiral type critical point. An accretion flow avoids to possess spiral type critical point as radial velocity gradient at this point becomes complex. The branches EF and GH intersect at a common point where the accretion rates at both the critical points are equal. Such a critical accretion rate is denoted by ${\dot {\cal M}_m}$ and the corresponding energy is represented by ${\cal E}_m$. Note that the flow with energy ${\cal E}>{\cal E}_{\rm G}$ has only one inner saddle type critical point. For ${\cal E}_m \le {\cal E} \le {\cal E}_{\rm G}$, the accretion rate ${\dot {\cal M}}_{\rm out} \ge {\dot {\cal M}}_{\rm in}$ and for ${\cal E}_{\rm H} \le {\cal E} \le {\cal E}_m$, ${\dot {\cal M}}_{\rm out} \le {\dot {\cal M}}_{\rm in}$, where the subscripts `$\rm in$' and `$\rm out$' denote the quantities at the inner and outer saddle type critical points respectively. Fig. \ref{E-mdot} clearly suggests that as $A$ is decreased, the overall range of EF and FG branches are reduced, whereas the range of GH branch is increased. Evidently, beyond a limiting value of $A$, the multiple critical points cease to exist and only the Bondi type accretion solutions passing through the outer critical points remains \cite{Chakrabarti:2004uy,Das-2007}.

\section{Accretion solutions in modified gravity}

In this section, we present the transonic global accretion solutions in modified gravity background. Moreover, we examine the role of gravity parameter ($A$) on the nature of the accretion solutions, and obtain the range of input parameters that admits transonic accretion solutions. 

\subsection{Global Transonic Solutions}

\begin{figure}
	\centering
	\includegraphics[width=\columnwidth]{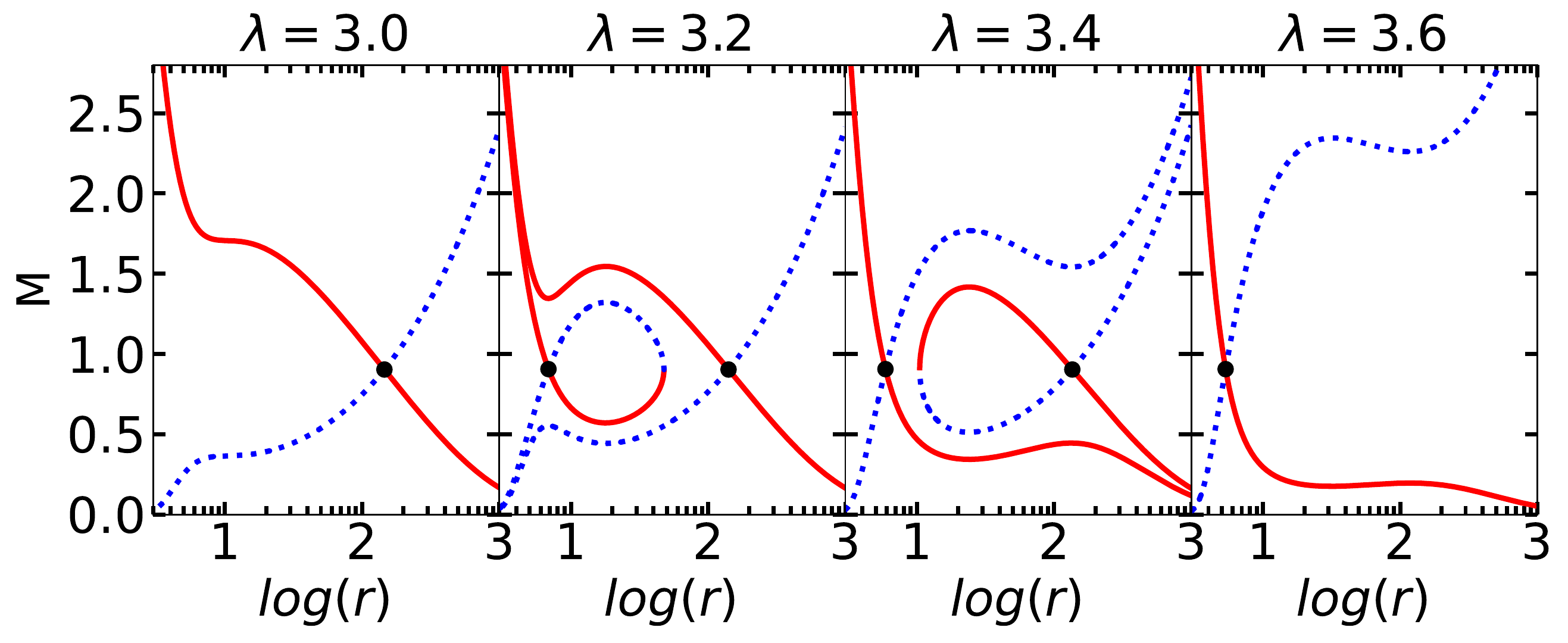}
	\bigbreak
	\includegraphics[width=\columnwidth]{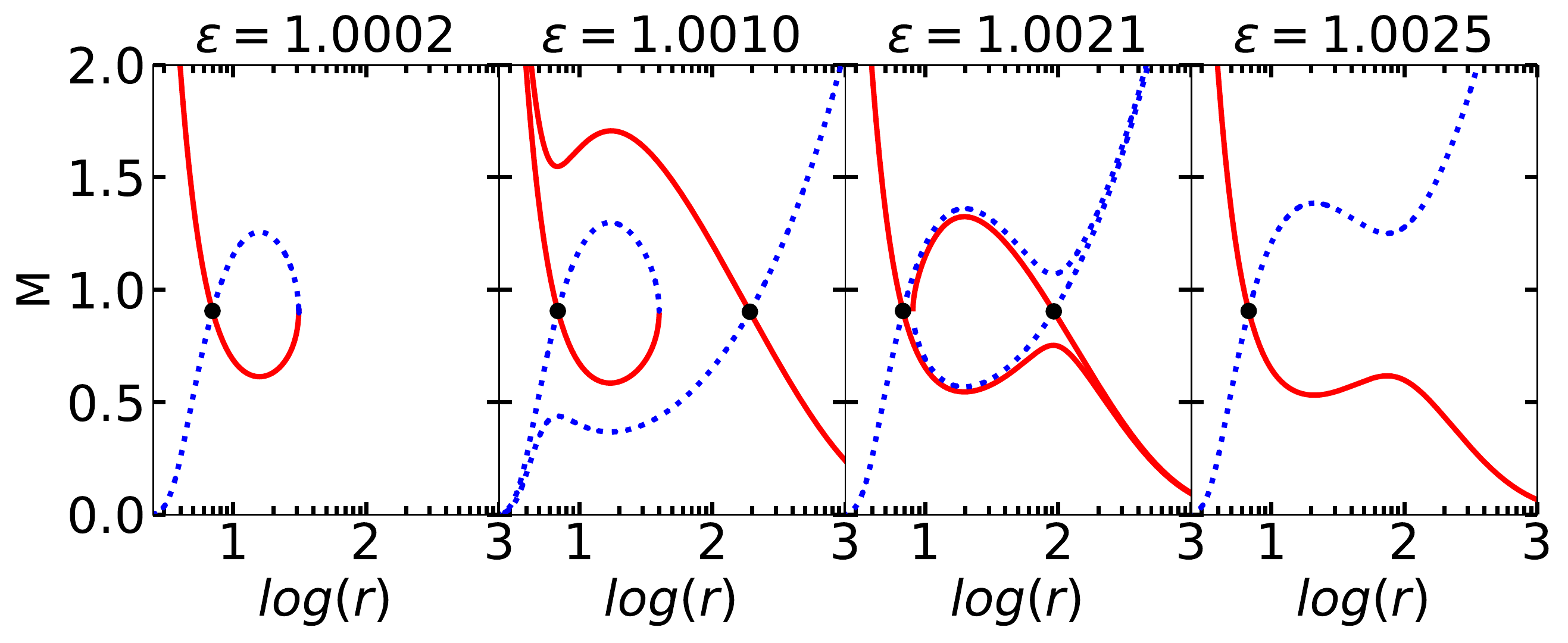}
	\caption{Variation of Mach number ($M=v/a_s$) as function of radial coordinate ($r$) for $A=-0.15$. Solid (red) curve denote accretion solution while dotted (blue) refers wind. {\it Top panels:} Here, ${\cal E}=1.0014$, and angular momentum is varied as $\lambda = 3.0,3.2,3.4$, and $3.6$, respectively. {\it Bottom panels:} Here, $\lambda = 3.2$ and energy is increased as ${\cal E}=1.0002, 1.0010, 1.0021$, and $1.0025$. See text for the details.
	}
	\label{M-r-L-E}
\end{figure}

Following \cite{Chakrabarti:2004uy}, we obtain the transonic global accretion solution by solving equations \eqref{e10}, \eqref{e17} and \eqref{e20}. In doing so, we use ${\cal E}$, $\lambda$, and $A$ as input parameters, and obtain $r_c$ and $(dv/dr)_c$ (see \$ III C). Employing these values, we integrate \eqref{e17} and \eqref{e20} from $r_c$ first inward up to horizon ($r_h$) and then outward towards the disk edge. Finally, we join both parts of the solution to obtain a global transonic accretion solution. In Fig. \ref{M-r-L-E}, we display the typical accretion solutions for different values of gravity parameter $A$, where Mach number ($M$) of the flow is plotted as function of radial distance ($r$). In the upper panel, we choose ${\cal E}=1.0014$ and $A=-0.15$, and gradually vary $\lambda$ from left to right. For $\lambda=3.0$, we find a single saddle type critical point (hereafter critical point) at $r_{\rm out}=145.3561$ and observe that transonic solution successfully connects the horizon ($r_h$) and outer edge of the disk ($r_{\rm edge}$) similar to Bondi solution \cite[]{Bindi-1952}. We refer this solution as global transonic solution which is of O-type \cite[]{Dihingia:2018tlr}. Here, critical point is marked by the filled circle (black), and solid (red) curve denotes the accretion solution, while the dotted (blue) curve is for wind. As the angular momentum of the flow is increased as $\lambda=3.2$, keeping the remaining parameter unchanged, we obtain multiple critical points, where inner critical point has appeared at $r_{\rm in}=6.8508$ along with the outer critical point at $r_{\rm out}=140.8278$. As before, the solution passing through $r_{\rm out}=140.8278$ yields as global solution, however, the solution passing through $r_{\rm in}=6.8508$ appears closed as it fails to connect $r_h$ and $r_{\rm edge}$. Interesting to note that accretion solution containing $r_{\rm in}=6.8508$ possesses higher entropy ($\dot {\cal M}$) compared to solution passing through $r_{\rm out}$, $i.e.$, $\dot {\cal M}_{\rm in} > \dot {\cal M}_{\rm out}$ (see Fig. \ref{E-mdot}). Solutions of these kinds are likely to be viable in the sense that they may trigger discontinuous transition of the flow variables in the form of shock waves \cite[and references therein]{Chakrabarti:1989,Chakrabarti:2004uy,Das-Chakrabarti2004,Das-2007,Sarkar-Das2016,Dihingia-etal2019a}. Indeed, the shock-induced global accretion solutions are potentially promising in the context of astrophysical application, namely, in explaining the timing as well as spectral features commonly observed from Galactic black hole sources \cite{Chakrabarti-Titarchuk1995,Mandal-Chakrabarti2005,Nandi-etal2012,Aktar-etal2017,Das-etal2021,Das-etal2022}. When angular momentum is increased further as $\lambda = 3.4$, the multiple critical points continue to exist, however, the overall characters of the accretion solution changes as the solution passing through $r_{\rm in}=5.8892$ opens up as it connects $r_h$ and $r_{\rm edge}$, whereas the solution passing through $r_{\rm out}=135.6049$ becomes closed with $\dot {\cal M}_{\rm in} < \dot {\cal M}_{\rm out}$. Global solutions passing through $r_{\rm in}$ are called as I-type \cite[]{Dihingia:2018tlr}. For $\lambda = 3.6$, we observe that $r_{\rm out}$ disappears and open solution passing through $r_{\rm in}=5.3357$ only remains. In the lower panels of Fig. \ref{M-r-L-E}, we choose $\lambda=3.2$ and $A=-0.15$, and vary ${\cal E}$. For ${\cal E}=1.0002$, we find that only inner critical point exists at $r_{\rm in}=6.9711$ and solution passing through it becomes closed. With the increase of energy as ${\cal E}=1.0010$, flow possesses multiple critical points at $r_{\rm in}=6.8890$ and $r_{\rm out}=191.6503$, where solution passing through $r_{\rm in}$ remain closed while open solution passes through $r_{\rm out}$. For ${\cal E}=1.0021$, multiple critical points continue to exist, however, the nature of the accretion solution alters. When energy is increased further as ${\cal E}=1.0025$, outer critical point ceases to exist although open solution passes through the inner critical point $r_{\rm in}=6.7540$. Overall, the above findings clearly indicates that both $\cal E$ and $\lambda$ seamlessly regulate the nature of the accretion solutions obtained from the modified gravity background.

\begin{figure}[h]
	\centering
	\includegraphics[width=\columnwidth]{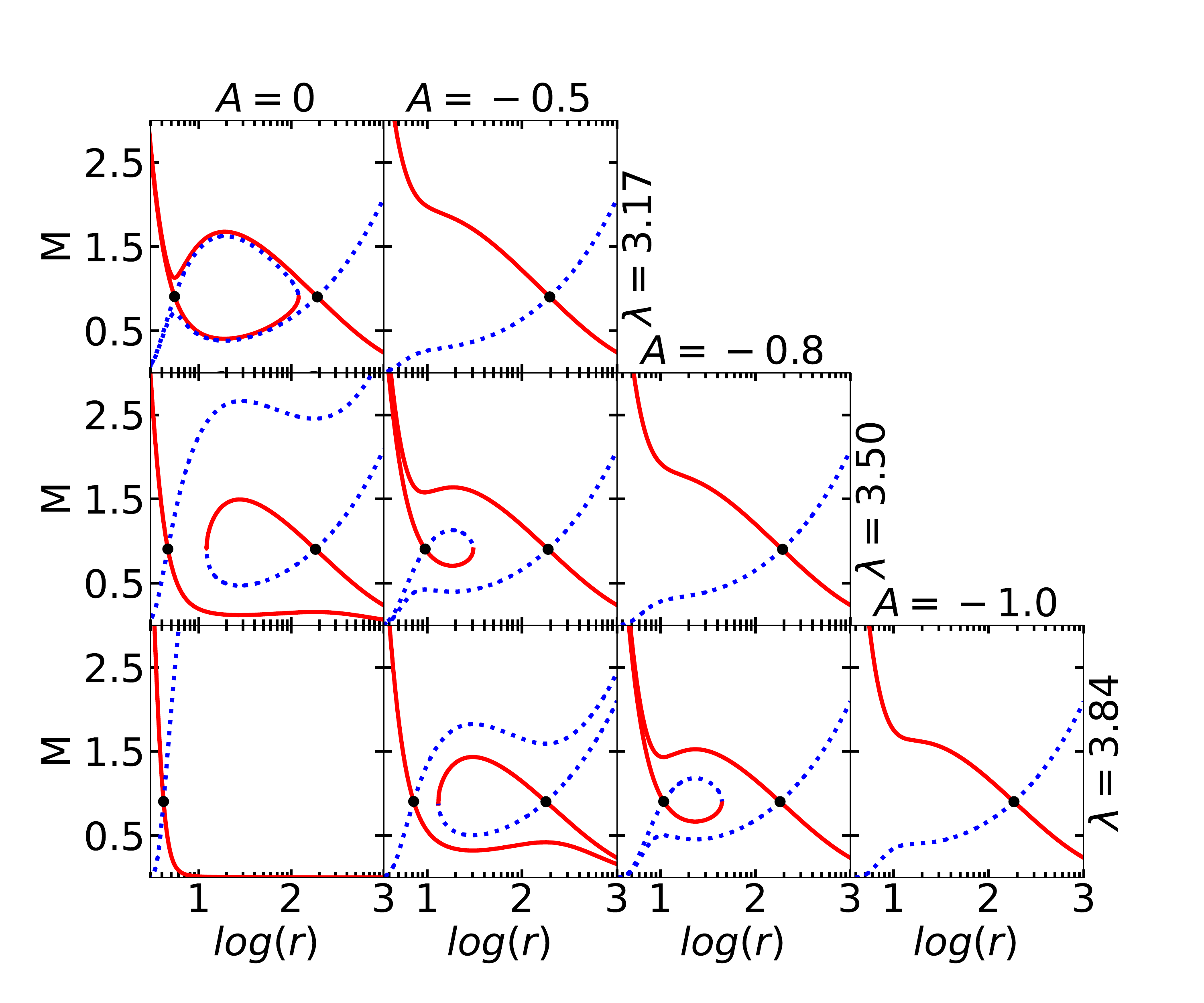}
	\caption{Modification of the accretion solution when the gravity parameter $A$ and angular momentum $\lambda$ are varied for flows with fixed energy ${\cal E}=1.001$. In each panel, solid (red) and dotted (blue) curves denote solutions corresponding to accretion and wind, respectively, and filled circles (black) refer to critical points. The chosen parameter values, such as $\lambda$ and $A$ are marked in each panel. See text for details.
	} 
	\label{M-r-A-L}
\end{figure}

In Fig. \ref{M-r-A-L}, we examine how gravity parameter $A$ alters the nature of the accretion solutions for flows with fixed energy ${\cal E}=1.001$. The obtained results are presented in each panel, where we show the variation of Mach number $M$ as function of $r$ for a given set of $\lambda$ and $A$ values. Here, solid (red) and dotted (blue) refer solutions corresponding to accretion and wind, and filled circles denote the critical point locations. For $\lambda=3.17$, we obtain two different types of solutions for $A=0$ and $-0.5$, as shown in the top panels. When angular momentum is increased as $\lambda=3.50$, three different types of solutions are found for $A=0$, $-0.5$, and $-0.8$ which are shown in the middle panels. For further increase of angular momentum as $\lambda=3.85$, four types of solutions are obtained for $A=0$, $-0.5$, $-0.8$, and $-1.0$, as shown in the bottom panels. Needless to mention that $A=0$ refers flow solutions corresponds to the Schwarzschild black hole.

\begin{figure}
	\centering
	\includegraphics[width=\columnwidth]{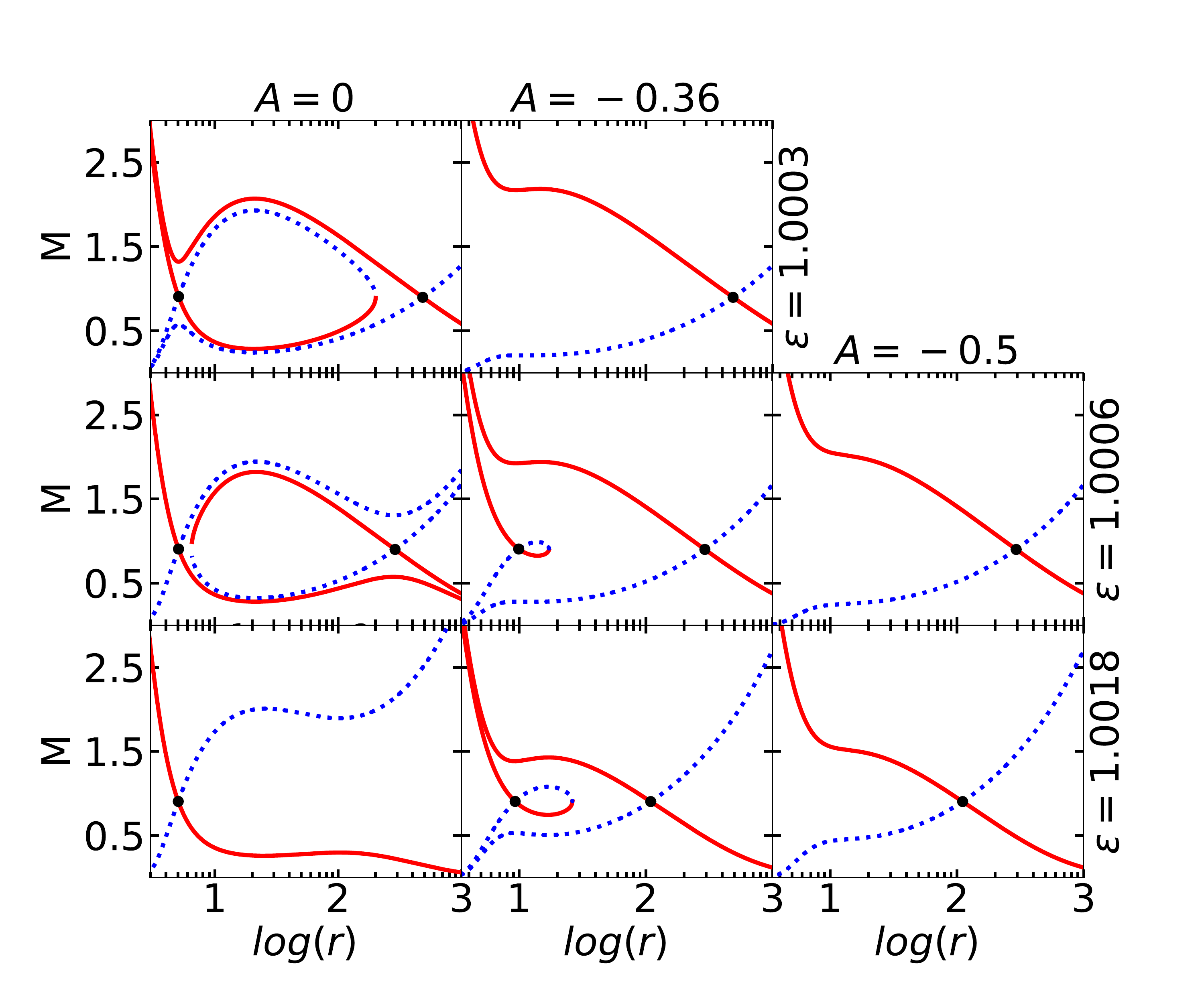}
	\caption{Same as Fig. \ref{M-r-A-L}, but $\cal E$ is varied keeping angular momentum fixed as ($\lambda=3.3$). See text for details.
	} 
	\label{M-r-A-E}
\end{figure}

We continue to study the combined role of gravity parameter ($A$) and flow energy ($\cal E$) in deciding the nature of the solutions for flows with fixed angular momentum as $\lambda=3.30$. In Fig. \ref{M-r-A-E}, we present the obtained results, where the chosen values of $A$ and ${\cal E}$ are marked in each panel. From the figure, it is evident that the gravity parameter ($A$) plays decisive role in deciding the overall character of the flow solutions in the frame work of modified gravity.

\begin{figure}
	\centering
	\includegraphics[width=\columnwidth]{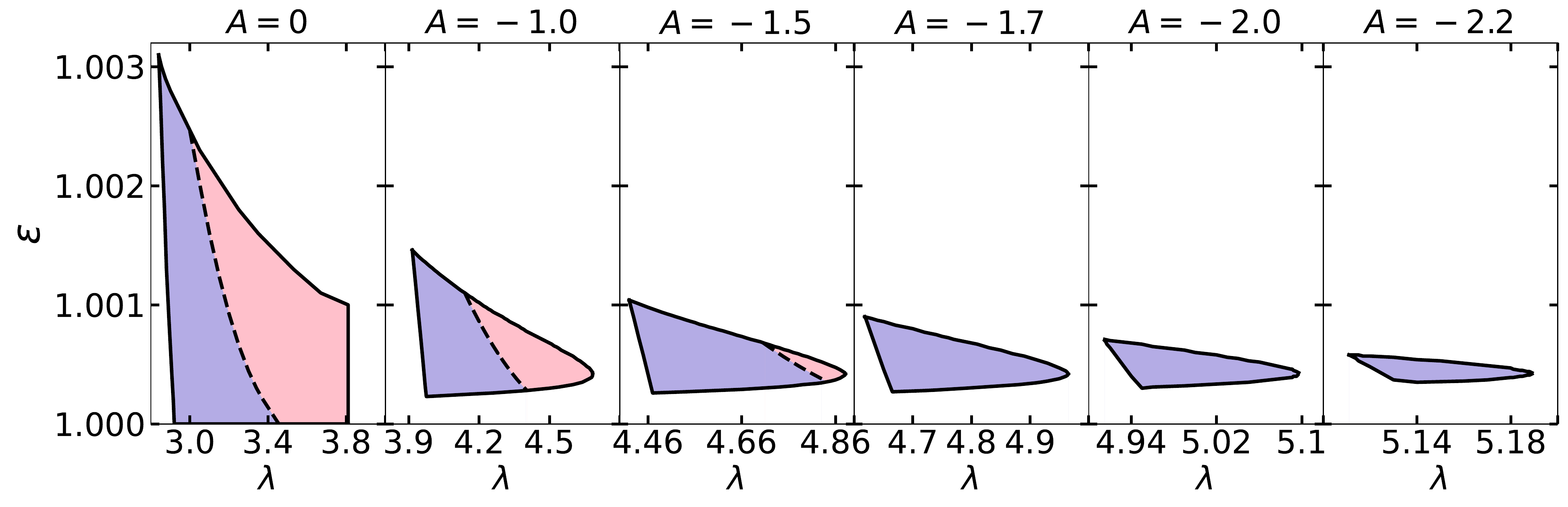}
	\caption{Modification of the parameter space in $\lambda-{\cal E}$ plane for multiple critical points. The region shaded in purple are for ${\dot{\cal M}}_{\rm in} > {\dot{\cal M}}_{\rm out}$, whereas orange shaded region accounts for ${\dot{\cal M}}_{\rm out} > {\dot{\cal M}}_{\rm in}$. The dashed line corresponds to ${\dot{\cal M}}_{\rm out} = {\dot{\cal M}}_{\rm in}$. In each panel, the gravity parameter $A$ is decreased from left to right which are marked. See text for details.
	}
	\label{L-E-A}
\end{figure}

\subsection{Parameter Space for multiple critical points}

Meanwhile, we indicate in \S III that depending on the input parameters (${\cal E}$, $\lambda$, $A$), accretion flow may possess either single or multiple critical points. It is also pointed out that in an accretion flow around black hole, the multiple critical points are essential for the triggering of discontinuous transition of the flow variables in the form of shock wave \cite[and references therein]{Chakrabarti:1989,Chakrabarti:2004uy,Das-Chakrabarti2004,Das-2007,Sarkar-Das2016,Dihingia-etal2019a}. Therefore, it is useful to identify the range of the input parameters that allows the flow to possess multiple critical points. Accordingly, in Fig. \ref{L-E-A}, we demonstrate the modification of the parameter space in $\lambda-{\cal E}$ plane with the variation of the gravity parameter $A$. In each panel, the domain enclosed by the solid (black) curve admits the multiple critical points and the same domain is further separated by the dashed (black) curve obtained for $\dot {\cal M}_{\rm in}=\dot {\cal M}_{\rm out}$. We observe that when $A$ is decreased, the effective domain of the parameter space is reduced with a marginal increase of minimum energy and shifted towards the higher angular momentum side. Moreover, we notice that the region shaded in orange is shrunk more than the purple shaded region and ultimately it ceases to exist leaving behind only the purple shaded region for $A=-1.667$. With the further decrease of the gravity parameter, the purple region completely disappears when $A$ goes down to $-2.34$.

\subsection{Radiative emission of the accretion flow}

The accretion flow around black hole generally becomes hot and dense because of the geometrical compression inevitable in the convergent accretion process. At this extreme environment, the flow remains in the plasma state and it is composed of both ions and electrons. Considering this, we infer that the flow is likely to radiate by means of free-free emissions \cite[]{Quenby2010}. The free-free emission rate per unit volume, per unit time and per unit frequency for bremsstrahlung process is given by,
\begin{equation} \label{e21}
    \epsilon(\nu) = \frac{32\pi e^6}{3m_e c^3} \left ( \frac{2\pi}{3k_B m_e T_e} \right )^{1/2} n^2_e \exp^{-h \nu /k_B T_e} g_{_{br}},
\end{equation}
where $e$, $m_e$ and $T_e$ are the charge, mass and temperature of electron, $k_B$ is the  Boltzmann constant, $h$ is the Planck’s constant, $\nu$ is the frequency, and $g_{\rm br}$ is Gaunt factor \cite{Karzas:1961}. In general, $g_{_{br}}$ assumes any values between $0.2$ to $5.5$ \cite{Karzas:1961}, and hence, we consider $g_{_{br}}=1$ in this work for simplicity. Following \cite{Chattopadhyay:2002ek}, we approximate the electron temperature as $T_e=\sqrt{m_e/m_p} T$, where $T$ refers the flow temperature. With this, we calculate the total luminosity emitted from the accretion disk as,
\begin{equation} \label{e22}
    L = 2 \int^{\infty}_{0} \int^{r_{\rm edge}}_{r_H} \int^{2\pi}_0 Hr\epsilon(\nu_e) d\nu_0 dr d\phi,
\end{equation}
where $\nu_e$ and $\nu_0$ denote the emitted and observed frequencies, respectively and are related as $\nu_e=(1+z)\nu_0$. Here $z$ is a red-shift factor and is given by \cite{Luminet:1979nyg},
\begin{equation} \label{e23}
    1+z = u^t(1+r \Omega \sin{\phi} \sin{i}),
\end{equation}
where $\Omega=u^\phi /u^t$ refers the angular velocity of the rotating flow and $i$ is the inclination angle of the observer with respect to the black hole. In this work, we choose $i=\pi/4$, mass of the black hole $M_S=10M_\odot$, $M_\odot$ being the solar mass and mass accretion rate $\dot{M}=0.1 \dot{M}_{Edd}$, where $\dot{M}_{\rm Edd}~\left[=1.44 \times 10^{17} \left(\frac{M_{S}}{M_\odot}\right)\right]$ g s$^{-1}$ is the Eddington accretion rate. 

\begin{figure}
	\centering
	\includegraphics[width=\columnwidth]{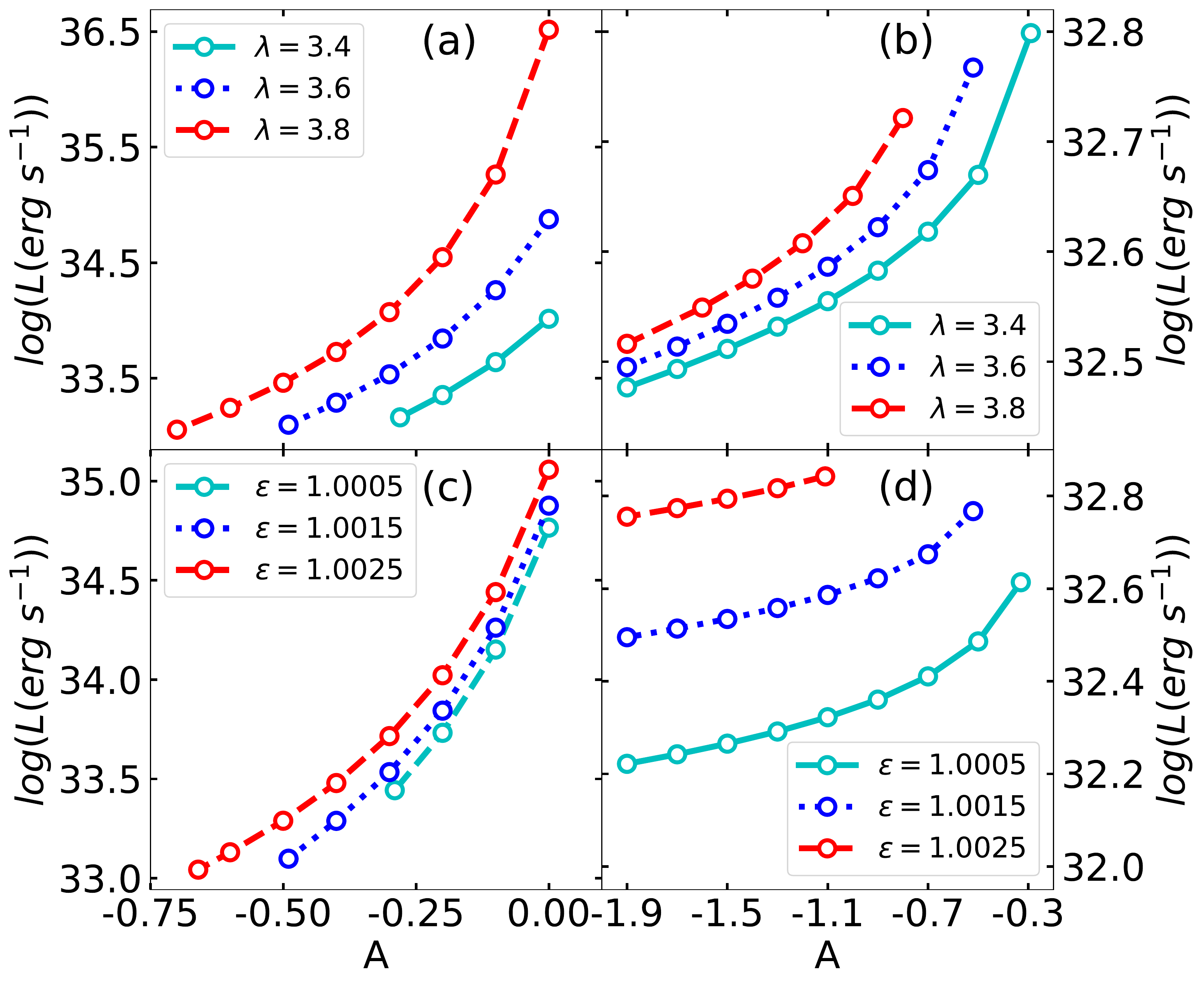}
	\caption{{\it Left:} Variation of disk luminosity ($L$) as function of gravity parameter $A$. In panel (a-b), we fix ${\cal E}=1.0015$ and open circles connected by solid (cyan), dotted (blue) and dashed (red) are for $\lambda = 3.4, 3.6$ and $3.8$, respectively. In panel (c-d), we choose $\lambda=3.6$ and open circles joined with solid (cyan), dotted (blue) and dashed (red) are for ${\cal E} = 1.0005, 1.0015$ and $1.0025$, respectively. See text for details. 
	} 
	\label{Lum-A}
\end{figure}

In Fig. \ref{Lum-A}, we demonstrate the variation of disk luminosity as function of gravity parameter ($A$). In the upper panels, we set the energy of the flow as ${\cal E}=1.0015$, and results are presented for different value of angular momentum ($\lambda$). In Fig. \ref{Lum-A}a, we focus only on those transonic global accretion solutions that passes through $r_{\rm in}$ (I-type solution), whereas results obtained from transonic global accretion solutions containing $r_{\rm out}$ (O-type solutions) are shown in Fig. \ref{Lum-A}b. In both panels, open circles join with solid (cyan), dotted (blue) and dashed (red) are for $\lambda = 3.4, 3.6$ and $3.8$, respectively. We observe that disk luminosity ($L$) increases with $A$ in both cases irrespective of the $\lambda$ values. Moreover, when $A$ is fixed, $L$ is seen to increase with the increase of $\lambda$. We also find that I-type solutions render luminosity variation relatively in the wider range in comparison to same obtained from O-type solutions. In the lower panels, we present the luminosity ($L$) variation with $A$ for a set of ${\cal E}$, where angular momentum is kept fixed at $\lambda=3.6$. We present the results obtained from I-type and O-type solutions in Fig. \ref{Lum-A}c and Fig. \ref{Lum-A}d, respectively. Here, open circles join with solid (cyan), dotted (blue) and dashed (red) are for ${\cal E} = 1.0005, 1.0015$ and $1.0025$, respectively. It is evident from the figure that flows having higher energies resulted enhanced luminosity as expected. What is more is that the maximum luminosity obtained from O-type global accretion solutions are generally remains smaller compared to the minimum luminosity obtained from the I-type global accretion solutions. This finding clearly indicates that I-Type solutions seem to be energetically more favourable over the O-type solutions in the frame work of modified gravity. 

\begin{figure}
	\centering
	\includegraphics[width=\columnwidth]{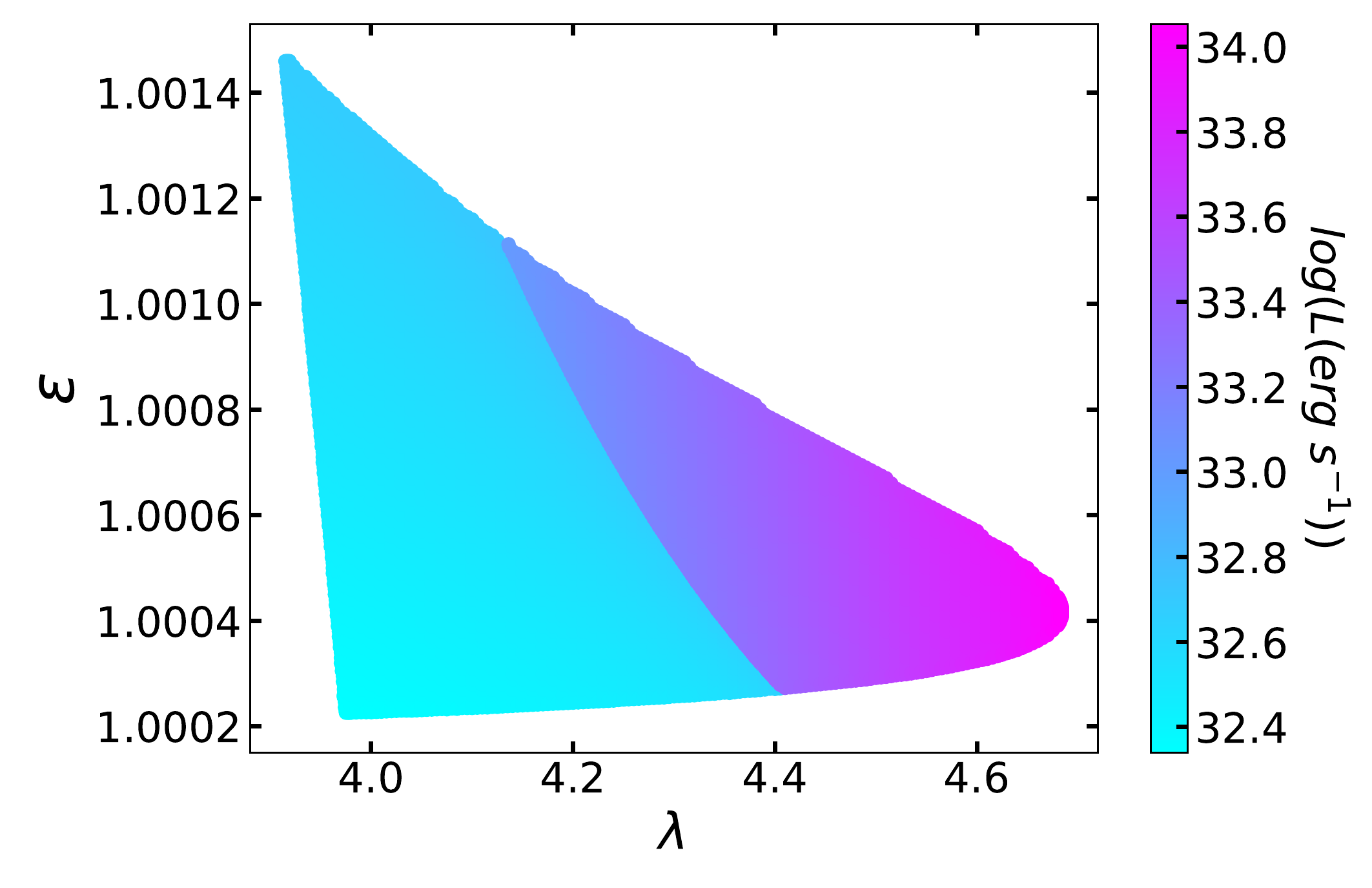}
	\caption{Two-dimensional surface projection of the three-dimensional plot of angular momentum ($\lambda$), energy (${\cal E}$) and disk luminosity ($L$), where multiple critical points are present. Here, we choose $A=-1.0$, and vertical colour coded bar represents the estimated range of disk luminosity ($L$). See text for details.
	}
	\label{L-E-Lum}
\end{figure}

For the purpose of completeness, in Fig. \ref{L-E-Lum}, we demonstrate the variation of disk luminosity ($L$) in $\lambda-{\cal E}$ plane, where two-dimensional projection of the three dimensional plot of $\lambda$, ${\cal E}$ and $L$ is depicted. Here, we keep the gravity parameter fixed as $A=-1.0$, and freely vary $\lambda$ and ${\cal E}$ in such a way that they render multiple critical points (both $r_{\rm in}$ and $r_{\rm out}$ simultaneously). We employ the global accretion solution passing through either $r_{\rm in}$ (I-type solution) or $r_{\rm out}$ (O-type solution) to calculate the disk luminosity ($L$) using equation (\ref{e22}). In the figure, vertical colour code indicates the range of disk luminosity as $2.18\times 10^{32}~{\rm erg}~{\rm s}^{-1} \le L \le 1.13 \times 10^{34} ~{\rm erg}~{\rm s}^{-1}$. We find that the luminosity is higher for the flows that pass through the inner critical point ($r_{\rm in}$) before connecting the horizon. Moreover, we observe the sub-division of the parameter space with sharp color contrast which is resulted due to the entropy separation as discussed \S IV.

\section{Conclusions}

In this work, we examine the properties of low angular momentum, advective accretion flows 
in the framework of $f(R)$ gravity. While doing this, we solve the governing equations that describe the flow motion in the steady state and for the first time to the best of our knowledge, obtain the global transonic accretion solutions in terms of the flow parameters, such as energy (${\cal E}$) and angular momentum ($\lambda$) for different values of the gravity parameter ($A$). The overall findings of this work are summarized below.

\begin{itemize}
	
	\item We find that the low angular momentum accretion flow becomes transonic before entering into the black hole in modified gravity model under consideration. Depending on the input parameters (${\cal E}, \lambda, A$), the accretion flow contains either single or multiple critical points (see Figs. \ref{E-r}-\ref{L-r}). Indeed, when the gravity parameter ($A$) is decreased, the possibility of obtaining the multiple critical points is decreased, and beyond its limiting value, multiple critical points disappear leaving the outer critical point only (see Fig. \ref{E-mdot}).
	
	\item With the suitable choice of the input parameters (${\cal E}, \lambda, A$), we obtain the global transonic accretion solution(s) passing through either single ($r_{\rm in}$ or $r_{\rm out}$) or multiple critical points ($r_{\rm in}$ and $r_{\rm out}$). Needless to mention that the overall nature of the accretion solution strictly depends on the input parameters (see Fig. \ref{M-r-L-E}). 
	
	\item One of the aims of the present paper is to study the effect of the gravity parameter $A$ on the nature of the global solutions. We find that for flows with fixed energy ${\cal E}$ and angular momentum $\lambda$, $A$ plays a pivotal role in deciding the nature of the accretion solution. When $A$ assumes relatively large negative values, we obtain accretion solutions similar to Bondi type irrespective of $\cal E$ and $\lambda$ values (see Figs. \ref{M-r-A-L}-\ref{M-r-A-E}).
	
	\item The comprehensive analysis of examining the nature of the critical points of the accretion flow in modified gravity suggests that a large region of the parameter space in $\lambda-{\cal E}$ plane provides multiple critical points. Accretion solutions containing multiple critical points with ${\dot {\cal M}}_{\rm in} > {\dot {\cal M}}_{\rm out}$ are potentially promising in explaining the timing and spectral features of Galactic black hole sources \cite{Chakrabarti-Titarchuk1995,Mandal-Chakrabarti2005,Nandi-etal2012,Aktar-etal2017,Das-etal2021,Das-etal2022}. What is more is that the effective domain of the parameter space is reduced with the decrease of $A$ values. We also notice that the parameter space with ${\dot {\cal M}}_{\rm out} > {\dot {\cal M}}_{\rm in}$ is susceptible to $A$ values as it disappears expeditiously as compared to the other part of the parameter space (see Fig. \ref{L-E-A}).
	
	\item We compute the disk luminosity ($L$) considering bremsstrahlung emission process. It is evident that $L$ strongly depends on the input parameters (${\cal E}, \lambda, A$). In fact, we find that for a fixed $A$, accretion solutions yield higher $L$ for flows with higher ${\cal E}$ and $\lambda$ (see Fig. \ref{Lum-A}). Further, we redraw the $\lambda-{\cal E}$ parameter space for multiple critical points to demonstrate the luminosity variations and indicate that the global accretion solutions containing $r_{\rm in}$ render higher disk luminosity $L$ (see Fig. \ref{L-E-Lum}).

\end{itemize}

Finally, we state the limitation of this work as it is developed based on several assumptions. We do not consider the effect of black hole rotation. We also ignore dissipative processes such as viscosity, magnetic fields etc., although they are expected to be relevant in the context of the accretion physics. Of course, the implementation of all these issues is beyond the scope of this paper which we plan to consider for future work and will be reported elsewhere.

\section*{Data Availability}

The data underlying this article will be available with reasonable request.

\section*{Acknowledgements}

Authors thank the anonymous reviewers for constructive comments and useful suggestions that help to improve the quality of the manuscript. Authors also thank the Department of Physics, Indian Institute of Technology Guwahati, for providing the facilities to complete this work. SD thanks Science and Engineering Research Board (SERB) of India for support under grant MTR/2020/000331. The work of SC is supported by the Science and Engineering Research Board (SERB) of India through grant MTR/2022/000318.

%


\vskip 0.2cm

\appendix

\section{Derivation of Innermost Stable Circular Orbits (ISCO)}
\label{Appendix A}

Consider a point particle moving in a circular orbit in a spherically symmetric space-time. The corresponding line element is given by ${\rm diag} (-s(r), p(r),r^2,r^2 \sin^2\theta)$. For circular orbit, the particle satisfies the equation given by,
	\begin{equation} \label{A1}
		V'_{\lambda_0}(r_0) = \frac{\partial V_\lambda(r)}{\partial r}\Bigr{|}_{r_0,\lambda_0}=0,
	\end{equation}
	where $V_\lambda(r)$ refer to the effective potential and $\lambda_0$ is the conserved orbital angular momentum for the circular orbit of radius $r_0$. Using the geodesics equations, we calculate  $\lambda_0$ as $\lambda_0=\sqrt{\frac{r_0^3 s^{\prime}(r_0)}{2 s(r_0)-s^{\prime}(r_0)r_0}}$. When family of circular orbit is considered, one can vary Eq. (\ref{A1}) to obtain 
	\begin{equation} \label{A2}
		\frac{\delta V'_{\lambda_0}(r_0)}{\delta r_0}=V''_{\lambda_0}(r_0)+\frac{\partial V'_{\lambda_0}(r_0)}{\partial \lambda_0} \frac{\delta \lambda_0}{\delta r_0}=0,
	\end{equation}
	where $\lambda_0$ is a function of $r_0$. Equation (\ref{A2}) yields as 
	\begin{equation} \label{A3}
		V''_{\lambda_0}(r_0)=-\frac{\partial V'_{\lambda_0}(r_0)}{\partial \lambda_0} \frac{\delta \lambda_0}{\delta r_0},
	\end{equation}
	where we assume $(\partial V'_{\lambda_0}(r_0)/\partial \lambda_0) \neq 0$ for $r_0 = r_{ISCO}$ and $\lambda_0=\lambda_{ISCO}$. Accordingly, the ISCO condition $V''_{\lambda_{ISCO}}(r_{ISCO})=0$ remains equivalent to $\delta \lambda_0/\delta r_0 = 0$ \cite[]{Song:2021ziq}.

\end{document}